# The Ground Truth Trade-Off in Wearable Sensing Studies

Daniyal Liaqat, BSc.[1, 2, 3], Robert Wu, MD, FRCPC, MSc.[4,1], Salaar Liaqat, MSc.[1], Eyal de Lara, PhD.[1], Andrea Gershon, MD, MSc.[5, 1], Frank Rudzicz, PhD.[6, 1, 7, 2]

[1]University of Toronto, [2]Vector Institute for Artificial Intelligence, [3]Tabiat Research, [4]University Health Network, [5]Sunnybrook Research Institute, [6]St Michael's Hospital, [7]Surgical Safety Technologies Inc

---

Perez *et al*'s study using the Apple Watch to identify atrial fibrillation (AF) is a watershed moment in large-scale machine learning for wearable computing. Identifying relevant patients will be tremendously important to research in healthcare. For a condition like AF, this could reduce stroke risk by two thirds.

Wearable sensing studies are interesting because they intersect two revolutionary technologies -- mobile computing and machine learning (ML). With 174 million smartwatches and 1.4 billion smartphones shipped in 2018 alone[1,2], these devices create a unique opportunity to collect sensor data, particularly for health monitoring. Advances in ML enable researchers to turn these data into meaningful insights about patients; however, there is a tradeoff between the quantity and the *quality* of the data that is often overlooked. Specifically, obtaining large quantities of *ground truth* or gold standard data in wearable studies remains a key challenge of the field.

In the study by Perez *et al*, only 450 out of 420,000 individuals had ground truth data. Their study excluded 417,000 participants using the irregular pulse notification. This design decision means their study was only able to report positive predictive value (PPV) and unable to explore sensitivity or specificity. In this editorial, we explore the difficulty of obtaining ground truth data and its implications for study design.

There remains uncertainty as to the true accuracy of mobile and wearable devices. For example, how the accuracy of measures such as steps or heart rate from smartwatches changes with various demographic covariates is an understudied question. The fundamental utility of consumer electronics in medicine has yet to be fully established, which is why we resort to gold standard devices, such as the ePatch used Perez *et al*'s study. Unfortunately, these devices are often burdensome; Perez *et al* received fewer ECG patches than anticipated. Additionally, deploying gold standard devices increases costs. Improvements to gold standard devices could make conducting studies easier. For participants, making these devices more comfortable and less prone to user error is and should continue to be a priority. To assist researchers, more gold standard devices should adopt standardized data formats, making obtaining and analyzing data easier. Ground truth devices enable studies like Perez *et al*'s that explores AF and Liaqat *et al*'s that explores respiratory rate using readily available consumer electronics. While these studies are interesting on their own, when their results are combined and applied to detecting higher level states such as acute exacerbations of COPD (Wu *et al*), they could bring about revolutionary changes to health management.

---

[1] https://www.statista.com/statistics/263437/global-smartphone-sales-to-end-users-since-2007
[2] https://www.statista.com/topics/4762/smartwatches/

These challenges with ground truth data are amplified when studying rare events. While wearable sensing studies collect vast amounts of data, only tiny fractions are relevant -- only 153 (0.0000036%) participants with AF were identified among 420,000. This means studies have to be sufficiently large or run long enough to capture enough events of interest. However, a larger or longer study means that the cost and user burden of ground truth devices becomes proportionally larger. Rare events of interest can also be a challenge for machine learning where many models are sensitive to class imbalances. Various approaches exist to the problem of imbalanced datasets, and should be applied in practice, including synthetic oversampling, training regimens such as mixup, and regularization.

One way to navigate these trade-offs is to be clear initially regarding the purpose of the study. A large consumer electronics company such as Apple may be interested in demonstrating that their devices do not cause unnecessary worry for users, nor burden to healthcare providers through false positives. For this objective, Perez *et al's* study was well designed, since PPV was important, but sensitivity and specificity were not. They demonstrated that of 420,000 users, only 2000 received an alert and, of those with valid ECG data, 34% had an irregular pulse, which is sufficient evidence that the AF detection on the Apple watch provides more benefit than harm. For a study exploring COPD exacerbations (Wu *et al*), where the objective is to equip COPD patients with a smartwatch that notifies them of exacerbations, low sensitivity would undermine the wearables key purpose. Limiting the population in the study to those with moderate to severe COPD who most need such an application increases the probability of observing exacerbations while reducing the overall study size. Furthermore, collecting ground truth data for only a small portion of the study can help overcome the cost and user burden challenges. These design decisions mitigate some of the challenges of ground truth data collection and enable reporting sensitivity.

# REFERENCES

1. Perez MV, Mahaffey KW, Hedlin H, et al. Large-Scale Assessment of a Smartwatch to Identify Atrial Fibrillation. *New England Journal of Medicine*. 2019;381(20):1909-1917. doi:[10.1056/NEJMoa1901183](10.1056/NEJMoa1901183)

2. Liaqat D, Abdalla M, Abed-Esfahani P, et al. WearBreathing: Real World Respiratory Rate Monitoring Using Smartwatches. *Proc ACM Interact Mob Wearable Ubiquitous Technol*. 2019;3(2):56:1–56:22. doi:[10.1145/3328927](10.1145/3328927)

3. Wu R, Liaqat D, de Lara E, et al. Feasibility of Using a Smartwatch to Intensively Monitor Patients With Chronic Obstructive Pulmonary Disease: Prospective Cohort Study. *JMIR mHealth and uHealth*. 2018;6(6):e10046. doi:[10.2196/10046](10.2196/10046)